\documentclass[prd,twocolumn,floatfix,preprintnumbers,letterpaper]{revtex4}
\usepackage{graphics}
\usepackage{epsfig}
\usepackage{amsmath}

\begin{document}

\title{Tracking quintessence and $k$-essence in a general cosmological background}
\author{Rupam Das} 
\author{Thomas W. Kephart}
\author{Robert J. Scherrer}
\affiliation{Department of Physics and Astronomy, Vanderbilt University,
Nashville, TN  ~~37235}

\begin{abstract}
We derive conditions for stable tracker solutions for both quintessence and
$k$-essence in a general cosmological background, $H^2 \propto f(\rho)$.
We find that tracker solutions are possible only when
$\eta \equiv d \ln f /d \ln \rho \approx constant$, aside from a few special
cases, which are enumerated.  Expressions for the quintessence
or $k$-essence equation of state are derived
as a function of $\eta$ and the equation of state of the dominant
background component.
\end{abstract}
\maketitle
%\baselineskip =24 pts
\section{Introduction}

The universe appears to consist of approximately 30\% nonrelativistic matter, including
both baryons and dark matter, and
70\% dark
energy (see Ref.
\cite{Sahni} for a recent review, and references therein).  The evolution of the
dark energy density depends on its equation of state, which
is usually parametrized in the form
\begin{equation}
p_{DE} = w \rho_{DE},
\end{equation}
where $p_{DE}$ and $\rho_{DE}$ are the pressure and density of the dark energy.
Then the density of the dark energy scales as
\begin{equation}
\label{rhoDE}
\rho_{DE} \propto R^{-3(1+w)}.
\end{equation}
The simplest model for the dark energy is a cosmological constant, for which $w = -1$ and
$\rho_{DE} = constant$.
More complex models have been proposed, in which the dark energy arises from a scalar field $\phi$; these
are called quintessence models \cite{ratra,turner,caldwelletal,liddle,Stein1}.  These models generally give rise to a time-varying
$w_\phi$ and more complex behavior for $\rho_{DE}$.  One advantage of such models is that certain classes of
quintessence potentials lead to tracker behavior, in which the evolution of the scalar field is
independent of the initial conditions.  The conditions for such tracking behavior have been worked out
in detail by Steinhardt, et al. \cite{Stein1}.

A second class of models generalizes quintessence to allow for a non-standard kinetic term.  These models,
dubbed $k$-essence, have also been explored in great detail
\cite{Arm1,Garriga,Chiba1,Arm2,Arm3,Chiba2,Chimento1,Chimento2,Scherrerk}.  These models can also lead to tracking
behavior, and the conditions necessary for such behavior have been discussed by Chiba \cite{Chiba2}.

Both quintessence and $k$-essence can be generalized to modified versions of the Friedmann equation.  In
the standard Friedmann equation, the relation between the scale factor $a$ (or, alternatively,
the Hubble parameter $H$) and the density is
\begin{equation}
H^2 = \left(\frac{\dot a}{a}\right)^2 = \frac{\rho}{3}.
\end{equation}
where we set $8 \pi G = 1$ throughout.
However, various proposals have been put forward to modify this equation at high energy.  In type II Randall-Sundrum
models, for example, one has \cite{RS,RS2}
\begin{equation}
H^2 \propto \rho^2,
\end{equation}
in the limit of large $\rho$,
while Gauss-Bonnet models can give \cite{GB}
\begin{equation}
H^2 \propto \rho^{2/3}.
\end{equation}
The Cardassian model \cite{Cardass} assumes an expansion law of the
form
\begin{equation}
\label{Cardeq}
H^2 = \frac{\rho}{3} + B \rho^n
\end{equation}
with $n < 2/3$.

Motivated by these examples, numerous authors
have examined the evolution
of various dark energy models in the context of
non-standard expansion laws \cite{Huey,Savchenko,Mizuno,Sami2,Copeland,Tsu1}.
The most general treatments are given in Refs. \cite{Sami2}-\cite{Tsu1}.
Sami et al. \cite{Sami2} examine quintessence
with constant $w_\phi$ for a power-law modification to the Friedmann equation
($H^2 \propto \rho^q$).  Copeland et al. \cite{Copeland} discuss ``scaling" quintessence
models, i.e., models for which $w_\phi = w_B$, with an arbitrary expansion
law, $H^2 \propto f(\rho)$.  Here $w_B$ is the ratio of pressure
to density for the dominant, ``background" fluid, e.g.,
$w_B = 0$ for a matter-dominated universe, and
$w_B = 1/3$ for a radiation-dominated universe.  Tsujikawa and Sami \cite{Tsu1} examine arbitrary scalar field
models (including both quintessence and $k$-essence) with scaling
behavior ($w_\phi = w_B$) in models with a power-law modification
to the Friedmann equation, $H^2 \propto \rho^q$. 

Here we generalize this earlier work by examining tracking solutions
for both quintessence and $k$-essence
in a general cosmological
background characterized by  ${H^{2}}\propto {f(\rho)}$.
Although we adopt the approach of Steinhardt et al. \cite{Stein1} for quintessence and Chiba \cite{Chiba2}
for $k$-essence, our formalism encompasses tracking solutions not only for a wide range of potentials but also for
a wide range of $f(\rho)$. We derive sufficient conditions for both $V(\phi)$ and $f(\rho)$ to obtain
tracking solutions with a constant $w_\phi$. This
formalism provides us with a generic method to study these solutions for a wide variety of
scalar field models such as quintessence, tachyon,
$k$-essence, and phantom models.

\section {Quintessence}

\subsection{Tracking solutions}

%We shall consider a scalar field with present equation-of-state
% $-1<w_Q<0$ in a flat
%cosmological background (consistent with inflation).  The ratio of the
%nergy density to the critical density
%today is $\Omega_Q$ for the $Q$-field and $\Omega_m$ for the baryonic
%and dark matter density where
%$\Omega_m + \Omega_Q=1$.  We use dimensionless units where the 
%Planck mass is $M_p=1$.%

The equation of motion for the $\phi$-field is
\begin{equation}
\label{motionq}
\ddot{\phi}+ 3H\dot{\phi} + V_\phi =0,
\end{equation}
where
\begin{equation}
V_\phi \equiv dV/d\phi,
\end{equation}
and
\begin{equation}
\label{H}
H^2 = \left(\frac{\dot{a}}{a}\right)^2 = f(\rho).
%{\beta}{{({\rho_{B}} + {\rho_%{\phi}}})}^ {n}.
\end{equation}
%\begin{equation}
%H^2 = \left(\frac{\dot{a}}{a}\right)^2=\kappa (\rho_m + \rho_r + \frac{1}{2} \dot{\phi}^2 + V)
%\end{equation}
\noindent Here $a$ is the Robertson-Walker scale factor,
and $\rho$ is the total density, given by
\begin{equation}
\rho = \rho_B + \rho_\phi,
\end{equation}
where
$\rho_B$ is the background (radiation + matter) density, and
$\rho_{\phi}$ is the scalar field energy
density.
The standard Hubble expansion law corresponds to equation (\ref{H}) with
$f(\rho)= \rho$; in this paper we allow $f(\rho)$ to have an arbitrary
functional form.
%$\beta$ is a constant. 

By definition, the tracking solutions are the solutions to which the evolution of the
scalar field $\phi$ converges for a wide range of initial conditions
for $\phi$ and $\dot{\phi}$.
We follow the approach
prescribed by Steinhardt et al. \cite{Stein1} for quintessence,
but now generalize it to the arbitrary expansion law given by equation (\ref{H}).
For tracking solutions,
$w_\phi$ is nearly constant \cite{Stein1}, where $w_\phi$ is given by 
\begin{equation} \label{qw}
w_{\phi} = \frac{p_\phi}{\rho_\phi} = \frac{\frac{1}{2}\dot{\phi}^2-
V}{\frac{1}{2}\dot{\phi}^2 + V}.
\end{equation}
It follows from equation (\ref{H}) that
\begin{equation}
\label{Hdot}
\dot{H} = {{\frac {3}{2}}{H^2}\eta{[({w_{\phi}-w_{B}})(1-{\Omega_{\phi}})-(1+w_{\phi})]}},
\end{equation}
where
$\eta$ encodes the information on the generalized expansion law in equation (\ref{H}):
\begin{equation}\label{eta}
\eta=\frac{d\ln f(\rho)}{d\ln(\rho)}.
\end{equation}
For the standard Hubble expansion, $\eta = 1$.  In this
paper, we will confine
our attention to the case $\eta > 0$, and our conclusions
will be valid only for this case.  However, we note that
$\eta < 0$ can lead to interesting types of behavior (e.g., a phantom-like
future singularity in a matter-dominated universe).  Several
specific models of this type are mentioned in Ref. \cite{Linder}.

By combining these relations, it is useful to cast the equation of motion into the following
form:
\begin{equation}\label{mq1}
 \frac{V_\phi}{\sqrt{V}} = \pm 3H\sqrt{\frac{1-w_{\phi}^2}{2}} (1+\frac{x'}{6}),
\end{equation}

%\begin{equation} \label{teq}
%\pm \frac{V'}{V} = 3 {\sqrt{\beta}}{{(\frac{1}{\Omega_{\phi}})}^{\frac{n}{2}}{%{\rho_{\phi}}^{\frac{n-1}{2}} \sqrt{1+w_{\phi}}
%\left[ 1+ \frac{1}{6} \frac{d\, \ln{x}}{d\,\ln{a}}\right],
%\end{equation}
\noindent where $x=(1+w_{\phi})/(1-w_{\phi})= \frac{1}{2}\dot{\phi}^2/V$ is the ratio of the
kinetic to potential energy for $\phi$, and 
${x'} \equiv d \,\ln{x}/d \, \ln{a}$. The $\pm$ sign depends on 
whether $V_\phi>0$ or $V_\phi<0$, respectively. It follows from equation (\ref{mq1}) that
the tracker condition ($\dot w_\phi\approx 0$) becomes
%The tracking solution (to which general solutions converge)
%has the property that $w_{\phi}$ is nearly constant and 
%lies between $w_B $ and $-1$.   For $1+w_{\phi} = {\cal O}(1)$,
%CHANGE
%$\dot{\phi}^2 \approx \Omega_{\phi} H^2$ and
%the  equation-of-motion, Eq.~(\ref{teq}),
%dictates that 
\begin{equation}\label{tc1}
\frac{V_\phi}{V^\frac {\eta+1}{2}} \approx \left(\frac{1}{\Omega_\phi}\right)^\frac{\eta}{2}.
\end{equation}
%for a tracking solution;
%we shall refer to this as the ``tracker condition."
This is the generalization of the Steinhardt et al. \cite{Stein1} tracking condition
to an arbitrary expansion law.

As in Ref. \cite{Stein1}, we define the function
\begin{equation}
\Gamma_V \equiv {V_{\phi\phi}}V/(V_\phi)^2,
\end{equation}
whose properties
determine whether tracking solutions exist. By taking the time derivative
of equation (\ref{mq1}) and combining with the
equation (\ref{Hdot}) and (\ref{mq1}) itself, we obtain the following equation:
\begin{eqnarray} \label{Gamma1}
&& \Gamma_V 
- {\frac {1+\eta}{2}} = \frac{\eta(w_B-w_{\phi}){\Omega_B}}{2(1+ w_{\phi})} \nonumber\\
&&
 - 
\frac{\eta(w_B-w_{\phi}){\Omega_B}+\eta+(\eta-2)w_{\phi}}{2(1+ w_{\phi})} \frac{{x'}}{6+{x'}}  \nonumber \\
&&
-\frac{2}{(1+w_{\phi})} \frac{{x''}}{(6+{x'})^2}.
\end{eqnarray}
where
${x''} \equiv d^2 \,\ln{x}/d \, \ln{a}^2$.
%Ivaylo: note that your X = (1/6) \dot{x}, as I understand it.-yes,it is
As expected, equation (\ref{Gamma1}) reduces to the corresponding
equation in Ref. \cite{Stein1} for $\eta=1$.
In a universe dominated by a background fluid ($\Omega_B \approx 1$) with
$w_\phi \approx$ constant and nearly
constant $\Gamma_V$, the above equation becomes
\begin{eqnarray}\label{ga}
\Gamma_V &\approx&  \frac {\eta+1}{2} +\frac{\eta(w_B-w_{\phi})}{2(1+
w_{\phi})},\nonumber\\
\label{tracking}
 &\approx& \frac{1}{2} + \frac{\eta}{2}\left(\frac{1+w_B}{1+w_\phi}\right).
\end{eqnarray}
In deriving the above equation, the plausibility of the condition that $\Gamma_V \approx$ constant has been discussed
in detail in Ref. \cite{Stein1}. The crucial point is that this condition
encompasses
a wide range of potentials including inverse power law potentials and combinations of inverse power law terms to give rise to tracking solutions.

We must know the appropriate restrictions on $\eta$,
i.e., on $f(\rho)$ to extract the tracking solutions from equation (\ref{Gamma1}).
Since the left-hand side of equation (\ref{ga}) is nearly constant,
it follows that $\eta$ must be nearly constant during
background domination, i.e., the function $f(\rho)$ must
satisfy (\ref{eta}) for a nearly constant $\eta$.
Thus we require an extra condition, in addition to
the conditions on $\Gamma_V$, to derive tracking solutions
for both quintessence and $k$-essence. It is
obvious that this extra condition arises from the
extra ``degree of freedom'' in choosing a different cosmological background.
The only
case for which $\eta$ is {\it exactly} constant is
$f(\rho)\propto \rho^n$ for a constant $n$.
This power-law behavior includes
both the Randall-Sundrum and Gauss-Bonnet models as special cases,
and it was studied in detail in Ref. \cite{Sami2}.
Of course, more general conditions can produce an expression
for $f(\rho)$ that is roughly constant over a wide range in the scale
factor.  For instance, a sum of power laws, e.g., as in equation (\ref{Cardeq}),
gives a value for $\eta$ that is nearly constant over most of the evolution of the universe, i.e.,
at all times except for the epoch when
the two contributions to $f(\rho)$ are roughly equal.

Note that there are a few trivial special cases for which this argument
breaks down.  In particular, if $V$ is a constant, the right hand
side of equation (\ref{mq1}) must be zero; this can be achieved
by taking $w_\phi = \pm 1$.  The case $w_\phi = -1$ corresponds
to a non-zero constant potential, while $w_\phi = 1$ is the solution
for $V=0$.  Both of these results are independent of the value of
$H$ on the right-hand side of equation (\ref{mq1}) and are
therefore independent of $\eta$.

The validity of equation (\ref{tracking})
may be checked by comparing with the results obtained by Sami et al.
\cite{Sami2}.
For scaling solutions with a constant $w_\phi$ in a background dominated universe,
the potential function takes the following form \cite{Sami2}
\begin{equation}\label{V}
V(\phi) \propto {\phi^{-\alpha}},
\end{equation}
where $\alpha$ is constant. Then we obtain from equation (\ref{ga})
\begin{equation}\label{ts1}
1+w_\phi \approx \eta(1+w_B)\frac {\alpha}{\alpha +2}.
\end{equation}
This solution agrees with the result obtained in Ref. \cite{Sami2}.

\subsection{Stability of the tracking solutions}
So far, we have derived solutions with constant
$w_\phi$ in a general cosmological background;
now we want
to check the stability of these solutions with constant $w_\phi$. 
In order to check the stability, we perturb the tracker value of $w_\phi$,
which we will call $w_0$, by an amount $\delta$. Then we
expand equation (\ref{Gamma1}) to
lowest order in $\delta$ and its derivatives to obtain
%\begin{equation}
%1+w_{\phi}={w_B-2(\Gamma_V -3/2)\over 2(\Gamma_V -3/2)+1}\simeq {\rm const}.\label%{eos}
%\end{equation}
\begin{eqnarray}\label{pq}
&&2\delta ^{''} + 3[\eta(1 + w_B) - 2w_0]\delta ^{'} \nonumber\\
&& + 9\eta (1+w_B)(1-w_0)\delta = 0,
\end{eqnarray}
where the prime means $d/d\ln a$ and $w_0$ is the value of $w_\phi$
derived from equation (\ref{ga}). The solution of this equation is
\begin{equation}\label{delta}
\delta \propto a^{\gamma},
\end{equation}
where
\begin{eqnarray}\label{gammap}
&&\gamma = -\frac{3}{4}\left[\eta(1+w_B) - 2w_0 \right]\nonumber\\
&&\pm \frac {3i}{4} \sqrt{{8\eta(1+w_B)(1-w_0)- \left[\eta(1+w_B)-2w_0\right]^{2}}}.\nonumber\\
\end{eqnarray}
In the derivation of this equation, $\Gamma_V$ and $\eta$ are
assumed to be constant.
 
In order to have $\delta$ decay, the real part of $\gamma$ has to be negative. Hence, it follows that
\begin{equation}\label{w0}
w_0 < \frac{\eta(1+w_B)}{2},
\end{equation}
provided the quantity under the square root is positive. If the quantity under the square root
is negative (so that both values are real), then the above equation is also a necessary condition since the first term under the square root is always positive, provided $\eta > 0$ and $w_0 < 1$. 
Using equation (\ref{ga}), the above inequality can be written in terms of $\Gamma_V$ as 
\begin{equation}\label{gammatracking}
\Gamma_V > \frac{3 \eta (1+w_B) + 2}{2 \eta (1+w_B) + 4}.
\end{equation}
Therefore, for a nearly constant $\Gamma_V$, $\eta$ and $w_\phi$, the tracker condition, i.e., equation (\ref{tc1}) gives the following possibilities:

{\bf a.} If $w_\phi < w_B$, then $\Omega_\phi$ increases with time.
Then we conclude from equation (\ref{tc1}) that $\vert {V_\phi}/{V^\frac{n+1}{2}}\vert$
decreases for a tracker solution. However, taking the time derivative of ${{V_\phi}/{V^{\frac{\eta+1}{2}}}}$, we obtain
\begin{equation}\label{tc3}
\frac{d}{dt}\left(\frac{V_\phi}{V^{\frac{\eta+1}{2}}}\right)={\frac{V_\phi^{2}}{V^{\frac{\eta+3}{2}}}}\dot{\phi}\left(\Gamma_V -{\frac{\eta+1}{2}}\right).
\end{equation}
Hence, $|{V_\phi}/{V^{\frac{\eta+1}{2}}}|$ decreases if  $\Gamma_V >{\frac{1+\eta}{2}}$.
Thus, $w_\phi < w_B$ is observed for 
\begin{equation}\label{tc4}
\Gamma_V >\frac{1+\eta}{2}.
\end{equation}
Combining this with the condition for stable tracking behavior
(equation \ref{gammatracking}), we obtain
\begin{equation}\label{gammatracking1}
\Gamma_V > max\left[\frac{3 \eta (1+w_B) + 2}
{2 \eta (1+w_B) + 4},\frac{1+\eta}{2}\right].
\end{equation}

This is the most interesting case, as it gives viable models
for an accelerating universe.
These conditions encompass more solutions than the ones derived in Refs. \cite{Sami2,Copeland,Tsu1}.
For example, for the exponential potential, we have $\Gamma_V =1$,
and the above conditions are satisfied as long as $\eta < 1$ (including, for example,
the Gauss-Bonnet expansion law).

{\bf b.} If $w_\phi > w_B$, then tracking behavior is observed for
\begin{equation}
\frac{3 \eta (1+w_B) + 2}{2 \eta (1+w_B) + 4}<\Gamma_V<\frac{\eta+1}{2}.
\end{equation}

{\bf c.}  If $\Gamma_V =({1+\eta})/{2}$, then $w_\phi=w_B$. This is one of the main results (using somewhat
different notation) derived in Ref. \cite{Copeland}.

\section{$k$-essence}
\subsection{Tracking solutions}

In general, $k$-essence can be defined as
any scalar field with non-canonical kinetic terms, but in practice such models are usually taken
to have a Lagrangian of the form:
\begin{equation}
\label{p}
{\cal L} = V(\phi)F(X),
\end{equation}
where $\phi$ is the scalar field, and $X$ is defined by
\begin{equation}
\label{grad}
X = \frac{1}{2} \nabla_\mu \phi \nabla^\mu \phi.
\end{equation}
The pressure in these models is given by
\begin{equation}
p_\phi = {\cal L},
\end{equation}
where ${\cal L}$ is given by
equation (\ref{p}), while the energy density is
\begin{equation}
\label{rho}
\rho_\phi = V(\phi)[2X F_X - F],
\end{equation}
where $F_X \equiv dF/dX$.
Therefore, the equation of state parameter, $w_\phi \equiv p_\phi/\rho_\phi$, is just
\begin{equation}
\label{w}
w_\phi = \frac{F}{2X F_X - F}.
\end{equation}

In defining the sound speed, we follow the convention of Garriga and Mukhanov \cite{Garriga}, who
argued that the relevant quantity for the growth of density perturbations is
\begin{equation}
\label{cs}
c_s^2 = \frac{(\partial p /\partial X)}{(\partial \rho/\partial X)} = 
\frac{F_X}{F_X + 2X F_{XX}},
\end{equation}
with $F_{XX} \equiv d^2 F/dX^2$.

In a flat Robertson-Walker metric, the equation of motion for the
$k$-essence field takes the
form:
\begin{equation}
\label{motion1}
(F_X + 2X F_{XX})\ddot \phi + 3H F_X \dot \phi + (2XF_X - F)\frac{V_\phi}{V} =
0.
\end{equation}
%\begin{equation}
%\label{H}
%H^2 = (\rho/3),
%\end{equation}
%and we take $8 \pi G =1$ throughout.  In equation (\ref{H}),
%$\rho$ is the total density, including the contributions of
%the background radiation and matter, as well as the $k$-essence
%component.
We can express the equation of motion for $\phi$ in an alternative form which
will be useful for subsequent analysis:
\begin{equation}\label{motion2}
\pm \frac{V_\phi}{V} \sqrt{2X} = H \left(\frac {1+w_\phi}{2}\right) (6+Ay'),
\end{equation}
where
%\begin{eqnarray}
%{V'\over V^{3/2}}&=&\mp {\kappa \over 2}
%\sqrt{{(1+w_{\p})F_X\over 3\Omega_{\p}} }
%\left(6+   A y'\right),\label{eom}\\
\begin{eqnarray}\label{A}
&&
A = \frac{(XF_X-F)(2XF_{XX}+F_X)}{ XF_X^2-FF_X-XFF_{XX}} \nonumber \\
&&
=\frac{1-w_{\phi}}{c_s^2-w_{\phi}},
\end{eqnarray}
 $y=(1+w_{\phi})/(1-w_{\phi})$ and 
$ y' =d\ln y/d\ln a$, and plus (minus) sign corresponds to $\dot\phi
<0~(\dot\phi>0)$, respectively. The tracker condition ($w_\phi \approx$ constant) becomes
\begin{equation}\label{tc2}
\pm \frac{V_\phi}{V^{(n+2)/2}} \approx \left(\frac{F}{\Omega_\phi}\right)^{{n}/{2}}\frac{1}{\sqrt{2X}}.
\end{equation}
It is not surprising to see that the tracker condition for $k$-essence has an extra ``degree of freedom'' in $F(X)$. The functional form of $F(X)$ plays a crucial role in determining the tracking conditions for $k$-essence and we shall consider it in the next section.

%\begin{equation}
%c_s^2={p_X\over p_X+2Xp_{XX}}.%={F_X\over W_X+2XW_{XX}}.
%\end{equation}
%Note that for quintessence with a canonical kinetic term, $c_s^2=1$. 
%For a tracker solution ($w_{\phi}\simeq $ const.),
%we obtain a relation:
%\begin{equation}
%{1\over \sqrt{\Omega_{\phi}}}=\mp {1\over \kappa\sqrt{3(1+w_{\phi})F_X}}
%{V'\over
%      V^{3/2}},\label{trac:condition}
%\end{equation}
%which might be called the k-essential counterpart of the tracker
%condition \cite{swz}. 

%Similar to quintessence, we define a dimensionless function $\Gamma_V$ 
%by  $\Gamma_\rho =VV''/V'^2$.
After taking 
the time derivative of equation (\ref{motion2}) and using equation (\ref{Hdot}), we obtain
\begin{eqnarray}\label{gakessence}
&&\Gamma_V - (1+\frac{\eta}{2})=\frac{\eta(w_B-w_\phi)\Omega_B}{2 (1+w_{\phi})}\nonumber\\
&&       -\frac{[\eta(w_B-w_\phi)\Omega_B +\eta+(\eta-2)w_\phi] Ay'}
{2(1+w_\phi)(6+Ay')}\nonumber\\
&&       -\frac{2(1-w_{\phi}) y''}{(1+w_\phi)(6+A y')^{2}(c_s^2-w_{\phi})} \nonumber\\
\label{kevol}
&&       -\frac{2\left(\dot w_{\phi}(1-c_s^2)-
(dc_s^2/dt)(1-w_{\phi})\right) y'/H} {(1+w_\phi)(6+A y')^{2}(c_s^2-w_{\phi})^2},
\end{eqnarray}
where $ y''=d^2\ln y/d\ln a^2$.  We note that for $\eta=1$, equation
(\ref{gakessence}) reduces to the one derived in Ref. \cite{Chiba2}.
%Also, it is interesting to note that the $RHS$ of equation (\ref{gakessence}) equals the $RHS$ of equation (\ref{Gamma1}) in the limit $c_s ^2 =1$ for canonical quintessence, though $w_\phi$ is different in each model. However, the equation of motion (\ref{motion1}) reduces to the following equation for $c_s ^2 =1$:
%\begin{equation}\label{reducedmotion}
%\ddot{\phi}+ 3H\dot{\phi} + \dot{\phi}^{2}\frac{V_\phi}{2V} =0.
%\end{equation}
%Obviously, equation (\ref{reducedmotion}) is different from (\ref{motionq}), and it is not surprising since the Lagrangian for quintessence is different from that for $k$-essence. Therefore, the difference in the $LHSs$ of equations (\ref{gakessence}) and (\ref{Gamma1}) implies that $k$-essence evolves differently than quintessence.
% Equation(\ref{tracker}) might be called 
%the k-essential counterpart of the tracker equation.
%Therefore, for the tracker solution (assuming $\Gamma_V \simeq$ const. 
%and $\Omega_{\phi}\ll 1$\footnote{This assumption is implicit in \cite{swz}.}) 
%we can write $w_{\phi}$ in terms of $\Gamma_V$:

For a background-dominated universe with a constant $w_\phi$ and almost constant $\Gamma_V$, the tracker equation (\ref{gakessence}) reduces to
\begin{eqnarray}
\Gamma_V &\approx& \frac {\eta+2}{2}  + \frac{\eta(w_{B} - w_{\phi})}{2(1+w_\phi)},\nonumber\\
\label{gamma2}
&\approx& 1 + \frac{\eta}{2}\left(\frac{1+w_B}{1+w_\phi}\right)
\end{eqnarray}
Note that equation (\ref{gamma2}) for $k$-essence
closely resembles equation (\ref{tracking}) for quintessence; the only difference
is the constant appearing in the first term.
For the standard Hubble expansion law ($\eta = 1$), we obtain
\begin{equation}
\Gamma_V \approx 1 + \frac{1}{2}\left(\frac{1+w_B}{1+w_\phi}\right),
\end{equation}
in agreement with the results of Ref. \cite{Chiba2}.

\subsection{Stability of the tracking solutions}

To determine the stability of the tracking solution, we
repeat the calculation of Sec. II.B. for the case of
$k$-essence.  We assume a $k$-essence field with equation of state
parameter $w_0$ and perturb $w_0$ by an amount $\delta$.  Then we expand
equation (\ref{kevol}) to lowest order in $\delta$ and its derivatives to obtain
\begin{eqnarray}\label{delta2}
&&2\delta ^{''} + 3[\eta(1+w_B)-2w_0]\delta ^{'} \nonumber\\
&& + 9\eta (1+w_B)(c_s^2-w_0)\delta = 0,
\end{eqnarray}
where the prime means $d/d\ln a$. The solution of this equation is
\begin{equation}
\delta \propto a^{\gamma},
\end{equation}
where
\begin{eqnarray}\label{gammap2}
&&\gamma = -\frac{3}{4}\left[\eta(1+w_B)-2w_0 \right]\nonumber\\
&&\pm \frac {3i}{4} \sqrt{8\eta(1+w_B)(c_s^2-w_0)-[\eta(1+w_B)-2w_0]^2}.\nonumber\\
\end{eqnarray}
Again, in order to have $\delta$ decay, the real part of $\gamma$ has to be negative. Hence, it follows that
\begin{equation}
\label{w01}
w_0 < \frac{\eta(1+w_B)}{2},
\end{equation}
and
\begin{equation}
\label{w02}
w_0 < c_s^2.
\end{equation}

At this point, the above conditions cannot be
translated into relations in terms of $\Gamma_V$
without considering the functional
form of $F(X)$, since $w_\phi$ and $c_s^2$
both depend on $F(X)$.
Now we discuss the restrictions on the
form of $F(X)$ for constant $w_\phi$.

A variety of functional forms for $F(X)$ and $V(\phi)$
have been considered in $k$-essence models (see, e.g.,
Refs. \cite{Arm2,Arm3}). However, we will focus on the form of
$F(X)$ responsible for stable tracking solutions for a constant
equation of state.
In order to find the functional form of $F(X)$ for stable tracking
solutions with constant $w_\phi$, we note that equation (\ref{w}) 
can be written as
\begin{equation}\label{lnf1}
{\frac{\partial {\ln F(X)}}{\partial {\ln X}}} = {\frac{1 + w_\phi}{ 2w_\phi}}.
\end{equation}

{\bf Case 1.}  The first possibility emerges
if we treat equation (\ref{lnf1}) as a differential equation and
derive the general solution, which is
\begin{equation}\label{xbeta}
F(X) = X^\beta, 
\end{equation}
where $\beta$ is a constant, and $w_\phi$ is then
\begin{equation}\label{beta}
w_\phi = \frac{1}{2 \beta - 1}.
\end{equation}
By inserting equation (\ref{xbeta}) into equation (\ref{cs}), we obtain 
\begin{equation}
c_s ^{2} = \frac{1}{{2\beta}- 1},
\end{equation}
so that
\begin{equation}
\label{wbeta}
c_s^2 = w_\phi.
\end{equation}
These solutions were previously derived
in Ref. \cite{Diez}; we note here that they are independent of $\eta$,
and therefore of the expansion law.  These solutions also do not
depend on the form of $V(\phi)$.

It is obvious from equation (\ref{wbeta}) that $c^{2}_{s}<0$ for any
of these models with negative pressure
($w_{\phi}<0$).  If
$c^{2}_{s}<0$, then the $k$-essence fluid is unstable
against perturbation.  Moreover, equation
(\ref{xbeta}) describes a phantom field for $0< \beta < 1/2$.

{\bf Case 2.}
A second class of solutions arises if the field evolves to a
state for which $X = X_0$, where $X_0$ is a constant \cite{Chimento1,Chiba1}.
In this case, we have \cite{Chiba1}
\begin{equation}
\label{lnf2}
\frac{\partial \ln F(X)} {{\partial \ln X}}\biggr|_{X=X_0} = \frac{1 + w_\phi}{ 2w_\phi}
\end{equation}
Again, we see that equation (\ref{lnf2}) is independent of $\eta$ and
hence, independent of the expansion law.  However, the condition for
a stable solution of the form
$X = X_0$ does depend on $\eta$, as we now show.

From equation (\ref{gamma2}),
the tracking conditions,
equations (\ref{w01})-(\ref{w02}), take the following form in terms of
$\Gamma_V$:
\begin{equation}\label{tckess1}
\Gamma_V > \frac{2\eta(1+w_B) + 2}{\eta(1+w_B) + 2},
\end{equation}
and
\begin{equation}\label{tckess2}
\Gamma_V > 1 + \frac{\eta(1+w_B)}{2(1+c_s^2)}.
\end{equation}
Therefore, for a nearly constant $\Gamma_V$, $\eta$, and $w_\phi$,
equation (\ref{tc2}) gives the following possibilities:

{\bf a.} If $w_\phi < w_B$, then $\Omega_\phi$ increases with time.
Then we conclude from equation (\ref{tc2}) that $\vert {\sqrt{2X}}{V_\phi}/{F^\frac{\eta}{2}}{V^\frac{\eta+2}{2}}\vert$ decreases for a tracker solution. However, taking
the time derivative of $\left( {\sqrt{2X}}{V_\phi}/{F^\frac{\eta}{2}}{V^\frac{\eta+2}{2}}\right)$, we obtain
\begin{equation}
\frac{d}{dt}\left(\frac{V_{\phi}}{V^{\frac{\eta+2}{2}}}{\frac{\sqrt {2X}}{F^\frac{\eta}{2}}}\right)={\frac{2X}{F^{\frac{\eta}{2}}}\frac{V_{\phi}^{2}}{V^{\frac{\eta+4}{2}}}}\left(\Gamma_V -{\frac{\eta+2}{2}}\right).
\end{equation}

In the derivation of this equation, we have used the condition that
$X=X_0$.
Hence,  $\vert {\sqrt{2X}}{V_\phi}/{F^\frac{\eta}{2}}{V^\frac{\eta+2}{2}}\vert$
decreases if  $\Gamma_V >(\eta+2)/2$. Thus, $w_\phi < w_B$ for 
\begin{equation}
\Gamma_V >\frac{\eta+2}{2}.
\end{equation} 
Combining this with the conditions
for stable tracking behavior (equations \ref{tckess1}-\ref{tckess2}), we obtain 
\begin{equation}
\Gamma_V > max[\frac{\eta+2}{2},\frac{2\eta(1+w_B) + 2}{\eta(1+w_B) + 2},1 +
\frac{\eta(1+w_B)}{2(1+c_s^2)}].
\end{equation}

{\bf b.} If $w_\phi > w_B$, then tracking behavior is observed for
\begin{equation}
max[\frac{2\eta(1+w_B) + 2}{\eta(1+w_B) + 2},1 + \frac{\eta(1+w_B)}{2(1+c_s^2)}]
< \Gamma_V < \frac{\eta+2}{2}.
\end{equation}

{\bf c.} If $\Gamma_V =({\eta+2})/{2}$, then $w_\phi=w_B$.
This case encompasses the solutions presented in Ref. \cite{Tsu1}.   

\section{DISCUSSION}

We have extended the formalism in Refs. \cite{Stein1} and \cite{Chiba2} to derive the tracker conditions for quintessence
and $k$-essence, respectively, for an arbitrary cosmological expansion law, $H^2 = f(\rho)$,
when the universe is dominated by a background fluid.  Our main
new result is that, with the exception of the special cases discussed above,
tracking solutions for either quintessence or $k$-essence are possible only
for $\eta = d \ln f /d \ln \rho \approx constant$, which is the case only when $f(\rho)$ is well-approximated
as a power-law.  In fact, such power-law
behavior corresponds to most of the models previously considered for non-standard expansion laws.

We note further that the expressions for $w_\phi$ for both quintessence and $k$-essence, and the conditions
for stable tracking behavior, can be derived by replacing $1+w_B$ with $\eta (1+w_B)$ in all of the corresponding
equations for the standard expansion law.  This is not surprising,
since a given value of $w_B$ corresponds to a background density scaling
as $\rho_B \propto a^{-3(1+w_B)}$.  Taking a constant value of $\eta$
in equation (\ref{H}) then gives $H^2 \propto a^{-3\eta(1+w_B)}$,
so $1+w_B$ is replaced by $\eta(1+w_B)$ in the expression for $H^2$
(see also the discussion in Ref. \cite{CLM}).

\acknowledgments

R.D., T.W.K., and R.J.S. were supported in part by the Department of Energy (DE-FG05-85ER40226).

\newcommand\AJ[3]{~Astron. J.{\bf ~#1}, #2~(#3)}
\newcommand\APJ[3]{~Astrophys. J.{\bf ~#1}, #2~ (#3)}
\newcommand\apjl[3]{~Astrophys. J. Lett. {\bf ~#1}, L#2~(#3)}
\newcommand\ass[3]{~Astrophys. Space Sci.{\bf ~#1}, #2~(#3)}
\newcommand\cqg[3]{~Class. Quant. Grav.{\bf ~#1}, #2~(#3)}
\newcommand\mnras[3]{~Mon. Not. R. Astron. Soc.{\bf ~#1}, #2~(#3)}
\newcommand\mpla[3]{~Mod. Phys. Lett. A{\bf ~#1}, #2~(#3)}
\newcommand\npb[3]{~Nucl. Phys. B{\bf ~#1}, #2~(#3)}
\newcommand\plb[3]{~Phys. Lett. B{\bf ~#1}, #2~(#3)}
\newcommand\pr[3]{~Phys. Rev.{\bf ~#1}, #2~(#3)}
\newcommand\PRL[3]{~Phys. Rev. Lett.{\bf ~#1}, #2~(#3)}
\newcommand\PRD[3]{~Phys. Rev. D{\bf ~#1}, #2~(#3)}
\newcommand\prog[3]{~Prog. Theor. Phys.{\bf ~#1}, #2~(#3)}
\newcommand\RMP[3]{~Rev. Mod. Phys.{\bf ~#1}, #2~(#3)}

\end{document}